\documentclass[11pt, a4paper, nofootinbib,aps,showpacs]{revtex4-1}

 \usepackage[bottom]{footmisc}

\usepackage[T1]{fontenc}
\usepackage[latin9]{inputenc}
\usepackage{mathrsfs}
\usepackage{bm}
\usepackage{amsmath}
\usepackage{amsthm}
\usepackage{amssymb}
\usepackage{stmaryrd}
\usepackage{mathtools}

\edef\ordinarycolon{\mathchar\the\mathcode`: }
\edef\ordinaryequals{\mathchar\the\mathcode`= }

\usepackage{breqn}
\catcode`^=7

\AtBeginDocument{%
  \catcode`^=12
  \def\coloneqq{%
    \mathrel{{\mathop\ordinarycolon}\mkern-1.2mu{\ordinaryequals}}%
  }%
}

\makeatletter

\allowdisplaybreaks

\let\cat@comma@active\@empty

\usepackage[capitalize]{cleveref}

\usepackage{color}
\RequirePackage[dvipsnames,usenames]{xcolor}

\DeclareMathOperator*{\argmin}{arg\,min}

\DeclareMathOperator*{\supp}{supp}

\newif\ifnotes
\notestrue

\ifnotes
 \newcommand{\dhwc}[1]{{\color{Red}{\bf{DHW COMMENT: #1}}}}   

 \newcommand{\ak}[1]{{\color{Orange}{{AK: #1}}}}
\else
 \newcommand{\dhwc}[1]{ }
 \newcommand{\dhws}[1]{ }
 \newcommand{\ak}[1]{ }
\fi

\newcommand{\todoPostArxiv}[1]{}

\renewcommand{\mid}{\vert}

\makeatother

\newcommand{\ba}{\begin{eqnarray}}
\newcommand{\ea}{\end{eqnarray}}
\newcommand{\eqn}[1]{\begin{align*}#1\end{align*}}
\newcommand{\eq}[1]{\begin{align}#1\end{align}}
\newcommand{\erf}[1]{Eq.~\eqref{#1}}

\newcommand{\p}{p}
\renewcommand{\r}{r}
\newcommand{\rr}{r}
\newcommand{\map}{\pi}
\newcommand{\Map}{\pi}

\newcommand{\Cin}{{\mathrm{IN}}}
\newcommand{\Cout}{{\mathrm{OUT}}}

\newcommand{\W}{{\mathcal{Q}}}

\newcommand{\EF}{{\mathcal{E}}}
\newcommand{\R}{{\mathbb{R}}}

\newcommand{\Z}{{\mathbb{Z}}}

\newcommand{\PP}{{\proc}}
\newcommand{\CC}{{C}}

\newcommand{\WWW}{{\mathcal{Q}}}
\newcommand{\WW}{{\mathcal{L}}}
\newcommand{\cs}[1]{\mathsf{#1}}   

\newcommand{\vl}{{\vec{\lambda}}}

\newcommand{\hn}{{{m}}}

\newcommand{\EPbase}{\sigma}
\newcommand{\EP}[1]{\EPbase_{#1}}
\newcommand{\EPmin}[1]{{\EPbase^{min}_{#1}}}

\newcommand{\DDbase}{D}
\newcommand{\KKbase}{S}
\newcommand{\IIbase}{\mathcal{I}}

\newcommand{\SSS}{{\mathcal{S}}}
\newcommand{\DD}{D}

\newcommand{\KKf}[2]{\KKbase(#1\Vert #2)}
\newcommand{\DDf}[2]{\DDbase(#1\Vert #2)}

\newcommand{\II}{\IIbase}

\newcommand{\ktlntwo}{k_B T \ln 2}

\newcommand{\pa}{\mathrm{pa}}

\newcommand{\proc}{}

\newtheorem{theorem}{Theorem}

\newtheorem{definition}{Definition}
\newtheorem{example}{Example}

\newtheorem{proposition}[theorem]{Proposition}

\begin{document}

\title{Overview of Information Theory, Computer Science Theory, and Stochastic Thermodynamics 
for Thermodynamics of Computation}

\author{David H. Wolpert}

\affiliation{Santa Fe Institute, Santa Fe, New Mexico \\
Arizona State University, Tempe, Arizona}

%
%

%
\begin{abstract}
\textbf{Abstract}: I give a quick overview of some of the theoretical
background necessary for using modern nonequilibrium statistical physics 
to investigate the thermodynamics of computation. I first present
some of the necessary concepts from information theory, and then
introduce some of the most important types of computational machine
considered in computer science theory. 

After this I present a central result from modern nonequilibrium statistical physics:
an exact expression for the entropy flow out of a system undergoing a given dynamics with
a given initial distribution over states. This central expression is crucial for analyzing how the total entropy flow 
out of a computer depends on its global structure, since that global structure
determines the initial distributions into all of the computer's subsystems,
and therefore (via the central expression) the entropy flows generated by all of those subsystems.
I illustrate these results by analyzing some of the subtleties concerning
the  benefits that are sometimes claimed for implementing an irreversible computation with
a reversible circuit constructed out of Fredkin gates.

\end{abstract}

\maketitle



\section{Introduction}

In this chapter I give a quick overview of some of the theoretical
background necessary for using modern nonequilibrium statistical physics 
to investigate the thermodynamics of computation.

I begin by presenting some general terminology, and then review some of the most directly relevant
concepts from information theory. Next I introduce
several of the most important kinds of computational machine studied in
computer science theory. After this I summarize the parts of nonequilibrium
statistical physics (more precisely, stochastic 
thermodynamics~\cite{seifert2012stochastic,van2015ensemble,esposito2010three}) that are
necessary for analyzing the thermodynamics of such computational machines.
The interested reader is directed to~\cite{wolpert_thermo_comp_review_2019} 
for a more detailed overview.

The key tool provided by stochastic thermodynamics is a 
decomposition of the entropy flow out of any open physical system that implements
some desired map $\map$ on some initial distribution $p_0$, into the sum of three 
terms~\cite{wolpert2018exact}: 
\begin{enumerate}
\item The ``Landauer cost'' of $\map$. This is an information-theoretic quantity:
the drop in Shannon entropy of the actual distribution over the states
of the system as the system evolves. The Landauer cost depends only on $\map$ and $p_0$,
being independent of the precise details of the physical system implementing $\map$.
\item The ``mismatch cost'' of implementing $\map$ with a particular
physical system. This is also an information-theoretic quantity:
a drop in Kullback-Leibler distance, between $p_0$ and a counterfactual ``prior'' distribution, $q_0$,
as those distributions both get transformed by $\map$. The mismatch cost depends only on $\map, p_0$, and $q_0$,
being independent of any details of the physical system implementing $\map$ that are not
captured in $q_0$.
\item The ``residual entropy production'' due to using a particular physical system. 
This is a \textit{linear} term, that depends on the graph-theoretic nature of $\map$, on $p_0$,
and on the precise details of the physical system implementing $\map$.
\end{enumerate}
\noindent (See~\cite{kolchinsky2016dependence,kolchinsky2017maximizing} for earlier
work that considered just the first two terms.) 

This decomposition allows
us to analyze how the thermodynamic costs of a fixed computational device vary
if we change the physical environment of that device, i.e., change 
the distribution of inputs to the device. This decomposition also plays
a key role in analyzing the thermodynamics of systems with multiple components
that are interconnected, since the physical environment of each of those components
is determined by the logical maps performed by the ``upstream'' components.
(See~\cite{wolpert2018exact,grochow_wolpert_sigact2018}.)

To illustrate this decomposition of entropy flow, 
I end by using it to analyze the relative thermodynamic
advantages and disadvantages of performing a computation with a logically reversible 
circuit (e.g. constructed out of Fredkin gates) rather than with a conventional circuit
that uses logically irreversible gates.

\section{Terminology and general notation}
\label{sec:notation}



As usual, for any set $A$, 
$A^{*}$ is the set of all finite strings of elements from $A$.
I write the Kronecker delta function as
\eq{
\delta(a, b) =
\begin{cases}
      & \text{1 if }a = b  \\
      & \text{0 otherwise}
\end{cases}
}
%
I write the indicator function for any Boolean function $f(z)$ as
\eq{
{\bf{I}}(f) =
\begin{cases}
      & \text{1 if }f(z) = 1   \\
      & \text{0 otherwise}
\end{cases}
}


In general, random variables are written with upper
case letters, and instances of those random variables are written with the corresponding lower case letters. When the context makes the meaning clear, I will
often also use the upper case letter indicating a random variable, e.g., $X$, 
to indicate the set of possible outcomes of that random variable.
For any distribution $\p(x)$ defined over the set $X$,
and any $X'\subseteq X$, I write $\p(X') = \sum_{x\in X'} \p(x)$.
Finally, given any conditional distribution $\map(y \mid x)$ and a distribution $\p$ over
$X$, I write $\map \p$ for the distribution over $Y$
induced by $\map$ and $\p$,
\eq{
(\map \p)(y) := \sum_{x\in X} \map(y\vert x) \p(x)
\label{eq:matrixnotation}
}


Computational machines are most often defined in terms of conditional 
distributions $\map$ that are (very good approximations of) single-valued state-update
functions. That means that there are transitions in the state of the system that cannot
occur in a single step, i.e. there are restrictions on which entries of
the transition matrix can be nonzero. In order to
analyze the entropy production in a physical system with this character,
it helps to ``decompose'' the dynamics of the system, into the entropy production
associated with a set of ``sub-maps'' defined by zero entries in the transition matrix.

Let $f$ be a single-valued
function from a set $X$ into itself. An \textbf{island}
of $f$ is a pre-image $f^{-1}(x)$ for some $x \in f(X)$~\cite{wolpert2018exact}.
I will write the set of all islands of a function $f$ as $L(f)$.

As an example, the logical AND operation,
\[
\map(c \vert a,b) = \delta(c, a \wedge b)
\]
has two islands, corresponding to $(a,b) \in \{\{0,0\},\allowbreak \{0,1\},\allowbreak \{1,0\}\}$ and $(a,b) \in \{\{1,1\} \}$, respectively.
I write the set of all distributions over an island $c \in L(f)$ as $\Delta_c$.
I make the obvious definitions that
for any distribution $p(x)$ and any $c \in L(f)$, the associated distribution
over islands is
$p(c)= \sum_{x\in c} p(x)$. As shorthand, I also write 
$p^c(x) = p(x\vert X\in c) = p(x){\bf{I}}(x \in c)/p(c)$.



Intuitively, the islands of a function
are different systems, isolated from one another for the duration
of any process that implements that function. 
As this suggests, the islands of a dynamic process depends on how long it runs. For example,
suppose $X = \{a, b, c\}$, 
and $f(a) = a, f(b) = a$, while $f(c) = b$. Then $f$ has two islands,
$\{a, b\}$ and $\{c\}$. However if we iterate $f$ we have
just a single island, since all three states get mapped under $f^2$ to the state $a$.

In the more general case where the physical process implements an arbitrary
stochastic matrix $\map$, the islands of $\map$ are 
the transitive closure of the equivalence relation,
\eq{
x \sim x' \Leftrightarrow \exists x'' : \map(x''|x) > 0, \map(x'' \mid x') > 0
}
%
(Note that $x$ and $x'$ may be in the same island even if there is no $x''$
such that both $P(x'' \mid x) > 0$ and $P(x' \mid x) > 0$, due to the transitive
closure requirement.) Equivalently, the islands of $\map$ are a partition $\{X^i\}$ of $X$
such that for all $X^i$, $x \not \in X^i$, 
\eq{
\supp \map(x' \mid x) \;\cap \bigcup_{x'' \in X^i} \supp \map(x' \mid x'') \;=\; \varnothing
}


Although the focus of this chapter is computers, which are typically viewed as
implementing single-valued functions, 
all of the results below also hold for physical processes that implement
general stochastic matrices as well, if one uses this more general definition
of islands. Note that for either single-valued or non-single-valued stochastic matrices $\map$,
the islands of $\pi$ will in general be a refinement of the islands of $\pi^2$, since
$\pi$ may map two of its islands to (separate) regions that mapped on top of one another by $\pi^2$.
This will not be the case though if $\map$ permutes its islands.


\section{Information theory}
\label{sec:info_notation}


The Shannon entropy of a distribution over a set $X$, 
the Kullback-Leibler (KL) divergence between two distributions
both defined over $X$ (sometimes called the ``relative entropy'' of those two distributions), 
and the cross-entropy between two such distributions, respectively, are defined as
\eq{
\SSS(\p(X)) &= -\sum_{x \in X} \p(x) \ln \p(x) \\
\DDf{\p(X)}{\r(X)} &= \sum_{x \in X} \p(x) \ln \frac{\p(x)}{\r(x)} \\
\KKf{\p(X)}{\r(X)} &= \SSS(\p(X)) + \DDf{\p(X)}{\r(X)} \label{eq:3} \\
& = -\sum_{x \in X} \p(x) \ln {\r(x)} 
}
(I adopt the convention of using natural logarithms rather than logarithms base 2
for most of this chapter.) I sometimes refer to the second arguments of KL
divergence and of cross-entropy as a \textbf{reference distribution}. Note that the
entropy of a distribution $p$ is just the negative of the KL divergence from $p$ to the uniform
reference distribution, up to an overall (negative) constant.

The conditional entropy of a random variable $X$ conditioned on 
a variable $Y$ under joint distribution $\p$ is defined as
\eq{
     \SSS(\p(X \mid Y)) &= \sum_{y \in Y} \p(y) \SSS(\p(X \mid y))  \nonumber \\
        &= -\sum_{x \in X, y \in Y} \p(y) \p(x \mid y) \ln \p(x \mid y)
}
and similarly for conditional KL divergence and conditional cross-entropy. 
The \textit{chain rule} for entropy~\cite{cover_elements_2012} says that
\eq{
\SSS(\p(X \mid Y)) + \SSS(\p(Y)) = \SSS(\p(X, Y))
}
Similarly,
given any two  distributions $p$ and $r$, both defined over $X \times Y$,
the conditional cross entropy between them 
equals the associated conditional entropy plus the associated conditional KL divergence:
\eq{
\KKf{\p(X \mid Y)}{\r(X \mid Y)} &= \SSS(\p(X \mid Y)) + \
    \DDf{\p(X \mid Y)}{\r(X\mid Y)}  \\
& = -\sum_{x \in X, y \in Y} \p(x, y) \ln {\r(x \mid y)} 
}

The mutual information between two random variables $X$ and $Y$ jointly
distributed according to $\p$ is defined as
\eq{
 I_\p(X ; Y) &\equiv \SSS(\p(X)) + \SSS(\p(Y)) - \SSS(\p(X, Y)) \\
 	&= \SSS(\p(X)) - \SSS(\p(X \mid Y))
\label{eq:mut_alt}	
}
(I drop the subscript $p$ where the distribution is clear from context.)
The \emph{data processing inequality} for mutual information~\cite{cover_elements_2012}
states that if we have random variables $X$, $Y$, and $Z$, and 
$Z$ is a stochastic function of $Y$, then $I(X;Z)\le I(X;Y)$.

Where the random variable is clear from context, I sometimes simply write $\SSS(\p)$, $\DDf{\p}{\r}$, and $\KKf{\p}{\r}$. I also sometimes abuse notation, and (for example) if $a$ and
$b$ are specified, write $\SSS(A = a \mid B =b)$ to
mean the conditional entropy of the random variable $\delta(A, a)$ conditioned on the 
event that the random variable $B$ has the value $b$.
When considering a set of random variables, I usually index them and their outcomes with subscripts, as in $X_1, X_2, \ldots$ and $x_1, x_2, \ldots$.  I also use notation like $X_{1,2}$ to indicate the joint random variable $(X_1,X_2)$.  



I write $S(\map \p)$ to refer to the entropy of distributions over $Y$ induced by $\p(x)$ and the conditional distribution $\map$, as defined in \erf{eq:matrixnotation}.
I use similar shorthand for the other information-theoretic quantities,
$\DDf{\cdot}{\cdot}$, $\KKf{\cdot}{\cdot}$ and $\II(\cdot)$.
In particular, the \emph{chain rule}  for KL divergence and the \emph{data-processing
inequality} for KL divergence, respectively, are~\cite{cover_elements_2012}:
\begin{enumerate}
\item  For all distributions $\p, \r$ over the space $X^a \times X^{b}$,
\eq{
&\DD\big(p(X^a, X^b) \mid\mid \r(X^a, X^b)\big) \nonumber \\
& \qquad\qquad =
   \DD\big( p(X^b) \mid\mid \r(X^b)\big) + \DD\big(p(X^a \mid X^b) \mid\mid \r(X^a \mid X^b)\big)
}
\item For all distributions $\p, \r$ over the space $X$ and conditional distributions $\map(y \mid x)$,
\eq{
\DDf{ \p }{ \r } &\ge \DDf{\map \p }{ \map \r}
}
\end{enumerate}
(Note that by combining the chain rule for KL divergence with the chain rule for entropy,
we get a chain rule for cross entropy.)

\section{Computational Machines}
\label{sec:comp_machines}

\subsection{Straight-line circuits}
\label{sec:circuit_theory}

A \textbf{(straight-line) circuit} $\CC$ is a tuple $(V, E, F, X)$ that can be
viewed as a special type of Bayes net~\cite{kofr09,ito2013information,ito_information_2015}. The pair $(V, E)$ specifies the vertices and edges of a directed acyclic graph (DAG). Intuitively, this DAG is the {wiring diagram} of the circuit.
(In the circuit engineering industry, this is sometimes called the ``netlist'' of the circuit.)
$X$ is a Cartesian product $\prod_v X_v$, where each $X_v$ is the space of the variable associated with node $v$.\footnote{In real electronic circuits, $x_v$ will typically
specify an  element in a coarse-graining of the microstate space of the system, but as mentioned above, that is not 
relevant here.} $F$ is a set of conditional distributions, one for
each non-root node $v$ of the DAG, mapping the joint value of the (variables at the) parents
of $v$, $x_{\pa(v)}$, to the value of (the variable at) $v$, $x_v$. In conventional circuit theory, $F$ is a set of single-valued functions. 

Note that following the convention in the Bayes nets literature, with this definition
of a circuit we orient edges in the direction
of logical implication, i.e., in the direction of information flow.
So the inputs to the circuit are the roots of the
associated DAG, and the outputs are the leaves. The reader should
be warned that this is the \textit{opposite} of the convention in computer science theory.
When there is no
risk of confusion, I simply refer to a circuit $\CC$, with all references to
$V, E, F$ or  $X$ implicitly assumed as the associated elements defining $\CC$.\footnote{More
general types of circuit than the one considered here allow branching conditions
at the nodes and / or loops among the nodes. Such circuits cannot be represented as a
Bayes net. To make clear what kind of circuit is being considered,
sometimes the branch-free, loop-free type of circuit is called a ``straight-line''
circuit~\cite{savage1998models}.} 

Straight line circuits are an example of \textbf{non-uniform} computers. These are
computers that can only work with inputs of some fixed
length. (In the case of a circuit, that length is specified
by the number of root nodes in the circuit's DAG.) One can use a single circuit to compute the output
for any input in a set of inputs, so long as all those inputs have the same length. If on the other hand 
one wishes to consider using circuits to compute the output for any input in a set that contains
inputs of all possible lengths, then one must use a \textbf{circuit family},
i.e., an infinite set of circuits $\{C_i : i \in \Z^+ \}$, where each circuit $C_i$ has
$i$ root nodes.

In contrast to non-uniform computers, \textbf{uniform} computers are machines that can work with arbitrary length
inputs.\footnote{The reader should be warned that computer scientists also consider ``uniform circuit
families'', which is something that is related but different.} 
In general, the number of iterations a particular uniform computational machine requires
to produce an output is not pre-fixed, in contrast to the case with any particular nonuniform computational machine. 
Indeed, for some inputs, a uniform computational machine may never
finish computing. The rest of this section introduces some of 
the more prominent uniform computational machines.

\subsection{Finite Automata}
\label{sec:sfa_def}

One important class of (uniform) computational machines are the finite automata.
There are several different, very similar definitions of finite
automata, some of which overlap with common definitions of
``finite state machines''. To fix the discussion, here I adopt the following definition:

\begin{definition}
\label{def:fa}
A \textbf{finite automaton} (FA) is a 5-tuple $(R, \Lambda, r^\varnothing, r^A, \rho)$ where:
\begin{enumerate}
\item $R$ is a finite set of \textbf{computational states};
\item $\Lambda$ is a finite (input) \textbf{alphabet};
\item $r^\varnothing \in R$ is the \textbf{start state};
\item $r^A \in R$ is the \textbf{accept state}; and
\item $\rho : R \times \Lambda \rightarrow R$ is the \textbf{update function},
mapping a current input symbol and the current computational state to a next
computational state.
\end{enumerate}
\end{definition}

A finite string of successive
input symbols, i.e., an element of $\Lambda^*$,
is sometimes called an \textbf{(input) word}, written as $\vl$. 
To operate a finite automaton on a particular input word, one
begins with the automaton in its start state, and feeds that state
together with the first symbol in the input word into the update function,
to produce a new computational state. Then one feeds in the next symbol
in the input word (if any), to produce a next computational state, and so on.
I will sometimes say that the \textbf{head} is in some state $r \in R$,
rather than say that the computational state of the automaton is $r$.

Often one is interested in whether the head is in state $r^A$
after the last symbol from the input word is processed. If
that is the case, one says that the automaton \textbf{accepts} that input word. In this
way any given automaton uniquely specifies a \textbf{language} of all input
words that that automaton accepts, which is called 
a \textbf{regular} language. As an example, any finite language (consisting
of a finite set of words) is a regular language. On the other hand,
the set of all palindromes over $\Lambda$, 
to give a simple example, is \textit{not} a regular language.

Importantly, any particular FA can process input words of arbitrary length.
This means that
one cannot model a given FA as some specific (and therefore fixed width) circuit, in general. 
The FA will have properties that are not captured by that circuit.
In this sense, individual FAs are computationally more powerful than individual circuits.
(This does not mean that individual FAs are more powerful than entire circuit \textit{families}
however; see the discussion in \cref{sec:TMs_def} of how circuit families can be even
more powerful than Turing machines.)

Finite automata play an important role in many different fields, including electrical
engineering, linguistics, computer science, philosophy, biology, mathematics, and logic. 
In computer science specifically, they are widely used in the
design of hardware digital systems, of compilers, of network protocols, and 
in the study of computation and languages more broadly.

To allow us to analyze a physical system that implements the running
of an FA on many successive input words, we need a way to signal to the system
when one input word ends and then another one begins. Accordingly, without loss of generality we assume
that $\Lambda$ contains a special \textbf{blank} state, written $b$, that delimits
the end of a word. I write $\Lambda_-$ for $\Lambda \setminus \{b\}$, so that
words are elements of $\Lambda_-^*$.

In a \textbf{stochastic} finite automaton (sometimes called a ``probabilistic
automaton''), the single-valued function $\rho$ is replaced by a conditional distribution. 
In order to use notation that covers all iterations $i$
of a stochastic finite automaton, I write this \textbf{update distribution} 
as $\Map(r_{i+1} \mid r_i, \lambda_i)$. The analogous extension of the word-based
definition of finite automata into a word-based definition of a stochastic
finite automata is immediate. For simplicity, from now on I will simply refer to ``finite
automaton'', using the acronym ``FA'', to refer to a finite automaton that
is either stochastic or deterministic.


Typically in the literature there is a set of multiple accept states --- called
``terminal states'', or ``accepting states'' --- not just one. Sometimes
there are also multiple start states.
%

\subsection{Transducers - Moore machines and Mealy machines}
\label{sec:transducer_def}
 
In the literature the definition of FA is sometimes extended 
so that in each transition from one computational state to the next an output symbol is 
generated.  Such systems are also called ``{transducers}'' in the computer
science community. 

\begin{definition}
\label{def:trans}
A \textbf{transducer} is a 6-tuple $(R, \Lambda, \Gamma, r^\varnothing, 
x^A, \rho)$ such that:

\begin{enumerate}
\item $R$ is a finite set, the set of \textbf{computational states};
\item $\Lambda$ is a finite set, called the \textbf{input alphabet};
\item $\Gamma$ is a finite set, called the \textbf{output alphabet};
\item $r^\varnothing \in R$ is the \textbf{start} state;
\item $r^A \in R$ is the \textbf{accept state};
\item $\rho: R \times {\Lambda} \rightarrow R \times {\Gamma}$
is the \textbf{update rule}.
\end{enumerate}
\end{definition}
\noindent
Sometimes the computational states of a transducer are referred to as
the states of its \textbf{head}. I refer to the (semifinite) string of symbols that have yet
to be processed by a transducer at some
moment as the \textbf{input (data) stream} at that moment. I refer to the string of symbols
that have already been produced by the information ratchet at some
moment as the \textbf{output (data) stream} at that moment.

To operate a transducer on a particular input data stream, one
begins with the machine in its start state, and feeds that state
together with the first symbol in the input stream into the update function,
to produce a new computational state, and a new output symbol. Then one feeds in the next symbol
in the input stream (if any), to produce a next computational state
and associated output symbol, and so on.

In a \textbf{stochastic} transducer, the single-valued function $\rho$ is replaced by
a conditional distribution. In order to use notation that covers all iterations $i$
of the transducer, I write this \textbf{update distribution} as
$\Map({\gamma}_{i+1}, r_{i+1} \mid r_{i},{\lambda}_i)$. 
%
Stochastic transducers are used in fields ranging from linguistics to
natural language processing (in particular machine translation) to
machine learning more broadly.
From now on I implicitly mean ``stochastic transducer'' when I use the
term ``transducer''.\footnote{The reader should be warned that
some of the literature refers to both FAs and transducers
as ``finite state machines'', using the term ``acceptor'' or ``recognizer''
to refer to the system defined in Def.~\ref{def:fa}. Similarly, the word
``transducer'' is sometimes used loosely in the physics community, to
apply to a specific system that transforms one variable --- typically
energy --- into another variable, or even just into another form.}

As with FAs, typically in the literature 
transducers are defined to have a set of multiple accept states, not just one. Sometimes
there are also multiple start states.
Similarly, in some of the literature
the transition function allows the transducer to receive
the empty string as an input and/or produce the empty string as
an output. 

A \textbf{Moore machine} is a transducer where the output ${\gamma}$ is
determined purely by the current state of the transducer, $r$. In contrast,
a transducer in which the output depends on both the current state $x$ and 
the current input ${\lambda}$ is called a \textbf{Mealy machine}.

As a final comment, an interesting variant of the transducers defined
in Def.~\ref{def:trans} arises if we remove the requirement that there be
accept states (and maybe even remove the requirement of start states). 
In this variant, rather than feeding an
infinite sequence of input words into the system, each of which results
in its own output word, one feeds in a single input word, which is
infinitely long, producing a single (infinitely long) output word. 
This variant is used to define so-called ``automata groups'' or ``self-similar 
groups''~\cite{MR2162164}.

Somewhat confusingly, although the computational properties of
this variant of transducers differs in crucial ways from those defined
in Def.~\ref{def:trans}, this variant is also called ``transducers''
in the literature on ``computational mechanics'', a branch of hidden Markov model 
theory~\cite{barnett2015computational}. Fortunately, 
this same variant have also been given a \textit{different} name, \textbf{information ratchets}, 
in work analyzing their statistical physics properties~\cite{mandal2012work}. 
Accordingly, here I adopt that term for this variant of (computer science)
transducers.

\subsection{Turing machines}
\label{sec:TMs_def}

Perhaps the most famous class of computational machines are Turing 
machines~\cite{hopcroft2000jd,arora2009computational,savage1998models}.
One reason for their fame is that it seems that one can model any computational machine that
is constructable by humans as a Turing machine. A bit more formally, the \textit{Church-Turing
thesis} states that, ``A function on the natural numbers is computable by a human 
being following an algorithm, ignoring resource limitations, if and only if 
it is computable by a Turing machine.'' 
The ``physical Church-Turing thesis'' modifies
that to say that the set of Turing machines includes 
all mechanical algorithmic procedures admissible by the laws of
physics.

In part due to this thesis,
Turing machines form one of the keystones of the entire field
of computer science theory, and in particular of computational complexity~\cite{moore2011nature}.
For example, the famous Clay prize question of whether $\cs{P} = \cs{NP}$ --- widely considered one of 
the deepest and most profound open questions in mathematics -- 
concerns the properties of Turing machines. As another example, the theory
of Turing machines is intimately related to deep results on the limitations of mathematics,
like G{\"o}del's incompleteness theorems, and has broader, philosophical
implications~\cite{aaronson2013philosophers}. As a result, it seems that the foundations
of physics may be restricted by some of the properties of Turing machines~\cite{barrow2011godel,aaro05}.

Along these lines, some authors have suggested that the foundations of statistical
physics should be modified to account for the properties of Turing machines, e.g.,
by adding terms to the definition of entropy. After all, given the
Church-Turing hypothesis, one might argue that the probability distributions at the heart of
statistical physics are distributions ``stored in the
mind'' of the human being analyzing a given statistical physical system (i.e., 
of a human being running a particular algorithm to compute a property of a given
system). See~\cite{caves1990entropy,caves1993information,zure89b}. 

There 
are many different definitions of Turing machines that are ``computationally equivalent''
to one another. This means that
any computation that can be done with one type of Turing machine can be done with the other.
It also means that the ``scaling function'' of one type of Turing machine, 
mapping the size of a computation to the minimal amount of resources needed to 
perform that computation by that type of Turing machine, is at most a polynomial
function of the scaling function of any other type of Turing machine. (See for example the
relation between the scaling functions of single-tape and multi-tape
Turing machines~\cite{arora2009computational}.) The following definition will be useful
for our purposes, even though it is more complicated than strictly needed:

\begin{definition}
A \textbf{Turing machine} (TM) is a 7-tuple $(R,\Lambda ,b,v,r^\varnothing,r^A,\rho)$ where:

%

\begin{enumerate}
\item $R$ is a finite set of \textbf{computational states};
\item $\Lambda$ is a finite \textbf{alphabet};
\item $b \in \Lambda$ is a special \textbf{blank} symbol;
\item $v \in \Z$ is a \textbf{pointer};
\item $r^\varnothing \in R$ is the \textbf{start state};
\item $r^A \in R$ is the \textbf{accept state}; and
\item $\rho : R \times \Z \times \Lambda^\infty \rightarrow 
R \times \Z \times \Lambda^\infty$ is the \textbf{update function}.
It is required that for all triples $(r, v, T)$, that if we write
$(r', v', T') = \rho(r, v, T)$, then $v'$ does not differ by more than $1$
from $v$, and the vector $T'$ is identical to the vectors $T$ for all components
with the possible exception of the component with index $v$;\footnote{Technically 
the update function only needs to be defined on the ``finitary'' subset of $\R \times \Z 
\times \Lambda^\infty$, namely, those elements of $\R \times \Z 
\times \Lambda^\infty$ for which the tape contents has a non-blank value in only finitely many positions.}
\end{enumerate}
\label{def:tm}
\end{definition}

$r^A$ is often called the ``halt state'' of the TM rather than the accept state. In addition,
in some alternative (computationally equivalent) definitions, there is a set of multiple accept states rather than
a single accept state. $\rho$ is sometimes called the ``transition function'' of the TM.
We sometimes refer to $R$ as the states of the ``head'' of the TM,
and refer to the third argument of $\rho$ as a \textbf{tape}, writing a
value of the tape (i.e., semi-infinite string of elements of the alphabet) as $T$.
The set of triples that are possible arguments to the update function 
of a given TM are sometimes called the set of \textbf{instantaneous descriptions}
(IDs) of the TM. (These are sometimes instead referred to as ``configurations''.)
Note that as an alternative to Def.~\ref{def:tm}, we
could define any TM as a map over an associated space of IDs.

Any TM $(R,\Sigma ,b,v,\rho ,r^\varnothing, \rho)$ starts with $r = r^\varnothing$, the counter
set to a specific initial value (e.g, $0$), and with $T$
consisting of a finite contiguous set of non-blank symbols, with
all other symbols equal to $b$. The TM operates by iteratively
applying $\rho$, until the computational state falls in $r^A$, at
which time it stops. (Note that the definition of $\rho$ for
$r = r^A$ is arbitrary and irrelevant.)

If running a TM on a given initial state of the tape results in the TM eventually halting,
the state of $T$ when it halts is called the TM's \textbf{output}. The initial
state of $T$ (excluding the blanks) is sometimes called the associated 
\textbf{input}, or \textbf{program}. (However,
the reader should be warned that the term ``program'' has been used by some physicists to
mean specifically the shortest input to a TM that results in it computing
a given output.) We also say that the TM \textbf{computes} an output
from an input.  In general, there will be inputs for which the TM never halts. 
The set of all those inputs to a TM that cause it to eventually
halt is called its \textbf{halting set}.

As mentioned, there are many variants of the definition of TMs provided above. In one
particularly popular variant the single tape in \cref{def:tm}
is replaced by multiple tapes. Typically one of
those tapes contains the input, one contains the TM's output (if and) when the TM
halts, and there are one or more intermediate ``work tapes'' that are
in essence used as scratch pads. The advantage of using this more complicated
variant of TMs is that it is often easier to prove theorems for such machines
than for single-tape TMs. However, there is no difference in
their computational power.

Returning to the TM variant defined in \cref{def:tm},
a \textbf{universal Turing machine} (UTM), $M$, is one that can be used
to emulate any other TM. More precisely, a UTM $M$ has the property that
for any other TM $M'$, there is an invertible map $f$ from the set of possible
states of the tape of $M'$ into the set of possible states of the tape of $M$, such
that if we:
\begin{enumerate}
\item  apply $f$ to an input string $\sigma'$ of $M'$ to fix an input string $\sigma$
of $M$; 
\item run $M$ on $\sigma$ until it halts; 
\item apply $f^{-1}$ to the resultant output of $M$;
\end{enumerate} 
then we get exactly the output computed by $M'$ if it is run directly on $\sigma'$.

An important theorem of computer science is that there exists universal TMs.
Intuitively, this just means that there exists programming languages which are ``universal'',
in that we can use them to implement any desired program in any other language, after
appropriate translation of that program from that other
language. This universality leads to a very important concept:

\begin{definition}
The \textbf{Kolmogorov complexity} of a UTM $M$ computing a string $\sigma \in \Lambda^*$
is the length of the shortest input string $s$ such that $M$ computes $\sigma$ from $s$.
\end{definition}
\noindent 
Intuitively, (output) strings that have low Kolmogorov complexity for some specific UTM $M$ are those with 
short, simple programs in the language of $M$. For example, in all common (universal) programming
languages (e.g., \textit{C, Python, Java}, etc.), 
the first $\hn$ digits of $\pi$ have low Kolmogorov complexity, since those
digits can be generated using a relatively short program.  
Strings that have high (Kolmogorov) complexity are sometimes referred to as
``incompressible''. These strings have no patterns in them that can be generated by
a simple program. As a result,
it is often argued that the expression ``random string'' should only be used for strings that 
are incompressible.

A \textbf{prefix (free) TM} is one such that no one input in its halting set
is a proper prefix of another string in its halting set (when both are viewed as symbol strings).
The \textbf{coin-flipping prior} of a prefix TM $M$ is the probability distribution 
over the strings in $M$'s halting set generated by IID ``tossing a coin'' 
to generate those strings, in a Bernoulli process, and then normalizing.\footnote{Kraft's 
inequality guarantees that since the set of strings in the halting set is a prefix-free
set, the sum over all its elements of their probabilities cannot exceed $1$, and
so can be normalized. See~\cite{livi08}.} So any string $\sigma$ in the halting set
has probability $2^{-|\sigma|} / \Omega$ under the coin-flipping prior, where
$\Omega$ is the normalization constant for the TM in question.

The coin-flipping prior provides a simple Bayesian interpretation of Kolmogorov
complexity: Under that
prior, the Kolmogorov complexity of any string $\sigma$ for any prefix TM $M$ is just
(the log of) the maximum a posterior (MAP) probability that any string $\sigma'$ in the halting
set of $M$ was the \textit{input} to $M$, conditioned on $\sigma$ being
the \textit{output} of that TM. (Strictly speaking, this result is only
true up to an additive constant, given by the log of the normalization
constant of the coin-flipping prior for $M$.)
%

The normalization constant $\Omega$ for any fixed prefix UTM, sometimes called ``Chaitin's Omega'',
has some extraordinary properties. For example, the successive digits of $\Omega$
provide the answers to \textit{all} well-posed mathematical problems. So if we
knew Chaitin's Omega for some particular prefix UTM, we could answer every problem in mathematics.
Alas, the value of $\Omega$ for any prefix UTM $M$ 
cannot be computed by any TM (either $M$ or some other one).
So under the Church-Turing hypothesis, we cannot calculate $\Omega$. 
(See also~\cite{baez2012algorithmic} for a discussion of a
``statistical physics'' interpretation of $\Omega$ that results if we view the coin-flipping prior
as a Boltzmann distribution for an appropriate Hamiltonian, 
so that $\Omega$ plays the role of a partition function.)

It is now conventional to analyze Kolmogorov complexity using prefix UTMs, with the coin-flipping
prior, since this removes some undesirable technical properties that Kolmogorov complexity has
for more general TMs and priors. Reflecting this, all analyses in the physics
community that concern TMs assume prefix UTMs. 
(See~\cite{livi08} for a discussion of related concepts like conditional Kolmogorov complexity.)

Interestingly, for all their computational power, there are some surprising ways
in which TMs are \textit{weaker} than the other computational machines introduced above.
For example, there are an infinite number of TMs that are more powerful than any given circuit, i.e., 
given any circuit $C$, there are an infinite number of TMs that compute the same function as $C$.
Indeed, any single UTM is more powerful than \textit{every} circuit in this sense. On the other hand,
it turns out that there are circuit \textit{families} that are more powerful than any single TM.
In particular, there are circuit families that can solve the halting problem~\cite{arora2009computational}.

\section{Entropy dynamics}
\label{sec:entropy_dynamics}

This section reviews those aspects of stochastic thermodynamics that are necessary
to analyze the dynamics of various types of entropy during the
evolution of computational machines. As illustrated with
examples, the familiar quantities at the heart of thermodynamics
(e.g., heat, dissipation, thermodynamic entropy, work) arise in special cases of this analysis.

In the first subsection, I review the conventional decomposition
of the entropy flow (EF) out of a physical system into the change in
entropy of that system plus the (irreversible) entropy creation (EP) produced
as that system evolves~\cite{van2015ensemble,seifert2012stochastic}. 
To fix details, I will concentrate on the 
total amount of EF, EP and entropy change that arise
over a time-interval $[0, 1]$.\footnote{In this paper
I will not specify units of time, and often implicitly change
them. For example, when analyzing the entropy dynamics of a given circuit,
sometimes the time interval $[0, 1]$ will refer to the time to 
run the entire circuit, and the attendant entropic costs.
However at other times $[0, 1]$ will refer to the time to run a single gate within
that circuit, and the entropic costs of running just that gate. In addition,
for computational machines that take more than one iteration to run, I will 
usually just refer
to a ``time interval $[0, 1]$'' without specifying which iteration of the machine 
that interval corresponds to.
Always the context will make the meaning clear.}

In the second subsection, I review recent results~\cite{wolpert2018exact} 
that specify how the EP generated by the evolution of some system depends on the initial distribution of states of the system.
These recent results allow us to evaluate how the EF of an arbitrary system, whose dynamics
implements some conditional distribution $\map$ of final states
given initial states, depends on the initial distribution of states of the system
that evolves according to $\map$. 
(As elaborated in subsequent sections, this dependence is one of the central features determining the
entropic costs of running any computational machine.)

I end this section with some general cautions about translating a computer science definition
of a computational machine into a physics definition of a system that implements that machine.

\subsection{Entropy flow, entropy production, and Landauer cost}
\label{sec:gen_land}

To make explicit connection with thermodynamics,
consider a physical system with countable state space $X$ that evolves over time 
interval $t\in [0,1]$ while in contact with one or more thermal reservoirs, while possibly also undergoing driving by 
one or more work reservoirs.\footnote{In statistical physics, a ``reservoir'' $R$ in contact 
with a system $S$ is loosely 
taken to mean an infinitely large system that interacts with $S$ on
time scales infinitely faster than the explicitly modeled dynamical evolution
of the state of $S$. For example, a ``particle reservoir'' exchanges particles with the
system, a ``thermal reservoir'' exchanges heat, and a ``work reservoir'' is an external system that
changes the energy spectrum of the system $S$.} In this chapter I focus on the scenario where 
the system dynamics over the time interval
is governed by a continuous-time Markov chain (CTMC). However many of the results 
presented below are more general.

Let $W_{x;x'}(t)$ be the rate matrix of the CTMC.
So the probability that the system is in state $x$ at time $t$ evolves according
to the linear, time-dependent equation
\eq{
\frac{d}{dt} p_x(t) = \sum_{x'} W_{x;x'}(t)p_{x'}(t)
\label{eq:rate_matix}
}
which I can write in vector form as $\dot{p}(t) = W(t) p(t)$. 
I just write ``$W$''
to refer to the entire time-history of the rate matrix.
$W$ and $p(0)$ jointly fix the conditional distribution of the system's
state at $t=1$ given its state at $t=0$, which I write as $\map$.
As shorthand, I sometimes abbreviate $x(0)$ as $x$, and sometimes abbreviate the initial distribution
$p(0)$ as $p$. (So for example, $\map p$ is the ending distribution over
states.) I will also sometimes abbreviate $p(1)$ as $p'$, and $x(1)$ as $x'$; the
context should always make the meaning clear. 

Next, define the {\textbf{entropy flow (rate)}}
at time $t$ as
\eq{
\sum_{x',x''} W_{x';x''}(t) p_{x''}(t) 
    \ln \bigg[\frac{W_{x';x''}}{W_{x'';x'}}\bigg]
\label{eq:EF_rate_def}
}
Physically, this corresponds to an entropy flow rate out of the system, into
reservoirs it is coupled to. 

In order to define an associated total amount of entropy flow 
during a non-infinitesimal time interval (EF), define $\bm{x} = (N, \vec{x}, \vec{\tau})$ 
as a trajectory of $N+1$ successive states $\vec{x} = (x(0), x(1), \dots, x(N))$, 
along with times $\vec{\tau} = (\tau_0 = 0, \tau_1, \tau_2, \dots, \tau_{N-1})$ of the
associated state transitions, where $\tau_{N-1} \le 1$, the time of the end of the process,
$x(0)$ is the beginning, $t=0$ state of the system, and $x(N)$ is the ending, $t=1$ state of the system.
Then under the dynamics of \cref{eq:rate_matix}, the probability of that 
$\bm{x}$ given the initial state $x_0$
is~\cite{esposito2010three,seifert2012stochastic,esposito2010threedetailed}
\begin{align}
\label{eq:trajweight}
p(\bm{x} \vert x(0)) =  \left(\prod_{i=1}^{N-1} S_{\tau_{i-1}}^{\tau_i}(x(i-1)) 
        W_{x(i); x(i-1)}(\tau_i) \right) S_{\tau_{N-1}}^1(x_{N})
\end{align}
where $S_{\tau}^{\tau'}(x) = e^{\int_{\tau}^{\tau'} W_{x; x}(t) dt}$ is the ``survival probability'' of remaining in state $x$ throughout the interval $t\in[\tau,\tau']$.
The total EF out of the system during the interval can be written as an integral
weighted by these probabilities:
\eq{
\WWW_\PP(p_0) &= \int  p_0(x_0) p(\bm{x} \vert x(0)) 
\sum_{i=1}^{N-1} W_{x(i); x(i-1)}(\tau_i)
\ln \frac{W_{x(i); x(i-1)}(\tau_i)}{W_{x(i-1); x(i)}(\tau_i)} \; D\bm{x} 
\label{eq:ctmc-ef}
}
%
(Note that I use the convention that EF reflects 
total entropy flow \textit{out} of the
system, whereas much of the literature defines EF as the entropy flow \textit{into} the 
system.) 

EF will be the central concern in the analysis below.
By plugging in the evolution equation for a CTMC, we can decompose 
EF as the sum of two terms. The first is just the change in entropy of the system during
the time interval. The second, is the \textbf{(total) entropy production} (EP)
in the system during the process~\cite{esposito2011second,seifert2012stochastic,van2015stochastic}.
I write EP as $\EP{} (p)$. It is the integral over the interval of the 
instantaneous EP rate,
\eq{
\sum_{x',x''} W_{x';x''}(t) p_{x''}(t)
   \ln \bigg[\frac{W_{x';x''}p_{x''}(t)}{W_{x'';x'}p_x'(t)}\bigg] 
\label{eq:ep_def}
}

I will use the expressions ``EF incurred by running a process'', 
``EF to run a process'', or
``EF generated by a process'' interchangeably, and similarly for EP.\footnote{Confusingly,
sometimes in the literature the term ``dissipation'' is used to refer to EP,
and sometimes it is used to refer to EF. Similarly, sometimes EP is instead referred to
as ``irreversible EP'', to contrast it with any change in the entropy of the
system that arises due to entropy flow.}
EF can be positive or negative. However, for any CTMC, the EP rate given in~\erf{eq:ep_def}
is non-negative~\cite{esposito2011second,seifert2012stochastic}, and therefore
so is the EP generated by the process. So
\ba
\WWW(p_0) &=& \EP{}(p_0) + S(p_0) - S(\map p_0) 
\label{eq:epv2}  \\
  &\ge& S(p_0) - S(\map p_0)
\label{eq:landauer_one_gate}
\ea
where throughout this section, 
$\map$ refers to the conditional distribution of the state of the system at $t = 1$
given its state at $t = 0$, which is implicitly fixed by $W(t)$.

Total entropy flow across a time interval can be written as
a linear function of the initial distribution: 
\eq{
\WWW_\PP(p_0) = \sum_{x_0} \mathcal{F}(x_0) p_0(x_0)
\label{eq:ef_linear}
}
for a function $\mathcal{F}(x)$ that depends on the entire
function $W_{x;x'}(t)$ for all $t \in [0, 1)$, and so
is related to the discrete time dynamics of the entire process, $\map(x_1 \mid x_0)$. (See~\erf{eq:ctmc-ef}.)
However, the \textit{minimal}
entropy flow for a fixed transition matrix $\map$ is the drop in entropy
from $S(p_0)$ to $S(\map P_0)$. This is not a linear
function of the initial distribution $p_0$. In addition, the entropy
production -- the difference between
actual entropy flow and minimal entropy flow -- is not a linear
function of $p_0$. These nonlinearities are the basis of much of the richness of statistical
physics, and in particular of its relation with information theory.

There are no temperatures in any of this analysis. Indeed, in this very
general setting, temperatures need not even be defined.
However, often the system is coupled to a heat bath,
with a well-defined temperature $T$.\footnote{Sometimes in the literature a ``heat bath''
is defined to be a thermal reservoir at (canonical ensemble) equilibrium, 
which is sometimes also presumed to be infinite. The context will make it clear 
whenever the discussion requires this presumption.} If in addition the Hamiltonian
of the system obeys ``detailed balance'' (DB) with that heat bath,
EF can be written as~\cite{esposito2010three}
\ba
\WWW = k_B T^{-1} Q
\label{eq:efheat}
\ea
where $k_B$ is Boltzmann constant, and $Q$ is the expected amount of heat transfered from the system into bath $\nu$ during the course of the process.

\begin{example}
Consider the special case of an \emph{isothermal} process, meaning there is a single heat bath at temperature $T$ (although possibly one or more work reservoirs
and particle reservoirs). Suppose that the process
transforms an initial distribution $p$ and Hamiltonian $H$ 
into a final distribution $p'$ and Hamiltonian $H'$. 
There are no \textit{a priori} requirements that either $p$ or $p'$ is at 
equilibrium for the associated Hamiltonian,
%

As mentioned, in this scenario 
EF equals $(k_B T)^{-1}$ times the total heat flow into the bath.
EP, on the other hand, equals $(k_B T)^{-1}$ times the \emph{dissipated work} of the process, which is the work done on the system over and above the minimal work required by any isothermal process that performs the transformation $(p,H)\mapsto (p',H')$~\cite{parrondo2015thermodynamics}. 
So by~\erf{eq:landauer_one_gate} and energy conservation, the minimal work is 
the change in the expected energy of the system plus ($k_B T$ times)
the drop in Shannon entropy of the system.
This is just the change in the \emph{nonequilibrium free energy} of the
system from the beginning to the end of the process~\cite{deffner2013information,parrondo2015thermodynamics,hasegawa2010generalization}.
\label{ex:single_bath}
\end{example}

There are many different physical phenomena that can result in nonzero EP. One
broad class of such phenomena arises if we take an ``inclusive'' approach,
modeling the dynamics of the system and bath together: 

\begin{example}
Continuing with the special case of an isothermal process,
suppose that the heat bath never
produces any entropy by itself, i.e., that the change in the entropy of the
bath equals the EF from the system into the bath. 
Then the change in the sum, \{marginal entropy of the system\}
$+$ \{marginal entropy of the heat bath\} must equal
the EF from the system to the bath plus the change in the marginal entropy of the system
by itself. By~\erf{eq:epv2} though, this is just the EP of the system

On the other hand, Liouville's theorem tells us that the \textit{joint} entropy of the 
system and the bath is constant. Combining establishes that EP of the system 
equals the change in the difference between the joint entropy and
the sum of the marginal entropies, i.e., EP equals the change in the
mutual information between the system and the bath.

To illustrate this, suppose we start with system and bath statistically independent. So
the mutual information between them
originally equals zero. Since mutual information cannot be negative, the change 
of that mutual information during the process is non-negative. This confirms that
EP is non-negative, for this particular case where we start with no statistical dependence
between the system and the bath. See~\cite{esposito2010entropy}.
\label{ex:3a}
\end{example}

Variants of \erf{eq:landauer_one_gate} are sometimes referred to in the literature
as the \textbf{generalized Landauer's bound}.
To motivate this name, suppose that there is a single heat bath, at
temperature $T$, and that the system has two possible states $X=\{0,1\}$. Suppose
further that the initial distribution $p(x)$ is uniform over these two states, and that the
conditional distribution $\map$ implements the function $\{0,1\} \mapsto 0$, i.e., a 2-to-1 `bit-erasure' map. So by \erf{eq:efheat} and the non-negativity of EP, the minimal heat flow {out}
of the system accompanying any process that performs that bit erasure is 
$k_B T (\ln 2 - \ln 1 ) = \ktlntwo$, in accord with the bound proposed by Landauer~\cite{landauer1961irreversibility}. 

Note though that in contrast to the bound proposed
by Landauer, the {generalized Landauer's bound} holds for systems with 
an arbitrary number of states, an arbitrary initial distribution over their states, and
an arbitrary conditional distribution $\map$. Most strikingly, the generalized Landauer bound
holds even if the system is coupled to multiple thermal reservoirs, all at
different temperatures, e.g., in a 
steady state heat engine~\cite{esposito2009thermoelectric,pietzonka2018universal}
(see Ex.~\ref{ex:multiple_baths} below). In such a case $\ktlntwo$
is not defined. Indeed, the generalized Landauer bound
holds even if the system does not obey detailed balance
with any of the one or more reservoirs it's coupled to.

Motivated by the generalized Landauer's bound, we define the \textbf{(unconstrained)
Landauer cost} as the minimal EF required to compute $\map$ on initial distribution
 $p$ using \textit{any} process, with no constraints:
\eq{
\WW(p, \map) := \SSS(p) - \SSS(\map p) \,.
\label{eq:land_cost_def}
}
With this definition we can write
\ba
\WWW(p) = \WW(p,\map) + \EP{\PP}(p)
\label{eq:efcombined}
\ea

\begin{example}
Landauer's bound is often stated in terms of the minimal amount of \emph{work} 
that must be done in order to perform a given computation, rather than the \emph{heat} that must be generated. 
This is appropriate for physical processes that both
begin and end with a constant, state-independent energy function.
For such processes,
there cannot be any change in expected energy between the beginning and end of the process. 
Moreover, by the first Law of thermodynamics, 
\[
\Delta E = W - \WWW(p) \,
\]
where $\Delta E$ is the change in expected energy from the beginning 
and end of the process, $W$ is work incurred by the process, and as before,
$\WWW_\PP(p)$ is the expected amount of heat that leaves the system and enters the bath.  
Since $\Delta E=0$, $W = Q$. So the bounds in \cref{ex:single_bath} 
on the minimal heat that must flow out of the system also give the
minimal work that must be done on the system.
\label{ex:3b}
\end{example}

Any process which achieves $\EP{}=0$ (i.e., the generalized Landauer's bound) for some 
particular initial distribution $p$ is said to be 
\textbf{thermodynamically reversible}, when run on that distribution. A necessary condition
for a process to be thermodynamically reversible is that if 
we run it forward on an initial distribution $p$ to produce $p'$, 
and then ``run the process backward'', by
changing the signs of all momenta and reversing the time-sequence of any driving,
we return to $p$.
(See~\cite{jarzynski_equalities_2011,van2015ensemble,sagawa2014thermodynamic,ouldridge_thermo_comp_book_2018}.)

A process being ``logically reversible'' means it implements
an invertible map over its state space. However, a process being 
``thermodynamically {reversible}'' does \textit{not} mean
it implements an {invertible} map over the space of all distributions. 
Indeed, unless a process implements a logically reversible map over
the state space, in general it will map multiple initial distributions
to the same final distribution, up to any desired accuracy~\cite{owen_number_2018}.

As an example, bit erasure is
a non-invertible map over the associated unit simplex. However,
for any initial distribution $q_0$, there
is a process that implements bit erasure on $q_0$ thermodynamically reversibly~\cite{esposito2010finite}.
Note though that if we run the bit erasure process
backwards from the ending (delta function) distribution,
we have to arrive back at the initial distribution $q_0$, to satisfy the
necessary condition for thermodynamic reversibility when run on $q_0$.
So if we run that bit-erasure
process on any initial distribution $p_0 \ne q_0$ and then run it
backwards, we would not arrive back at $p_0$ (we arrive at $q_0$ instead). 
This proves that the bit-erasure process cannot
be thermodynamically reversible when run on any such $p_0 \ne q_0$.

This  bit-erasure example underscores that thermodynamic
reversibility is a joint property of a process $W$ and the initial distribution $p_0$;
if we run the same process on a different initial distribution, in general
the amount of EP it generates will change. This dependence of EP on the initial distribution
is a central issue in analyzing the entropic costs of computation, and is addressed
in the next subsection.

\subsection{Mismatch cost and residual EP}

\label{sec:motivation}

Computational machines are built of multiple interconnected computational devices.
A crucial concern in calculating the entropic costs of running such a computational machine
is how the costs incurred by running any of its component devices, implementing some distribution $\map$,
depends on the distribution over the inputs to that device, $p_0$.
It is crucial how the entropic cost of running an AND gate depends on its inputs, how the cost of running an OR gate depends on its inputs, etc.

For a fixed $\map$, we 
can write Landauer cost of any process that implements $\map$
as a single-valued function of the initial distribution
$p_0$; no properties of the rate matrix $W$ matter for calculating Landauer cost, beyond the fact that
that matrix implements $\map$. However, even if we fix $\map$, we cannot write EP as a single-valued
function of $p_0$, because EP \textit{does} depend on the details of how $W$ implements $\map$. 
(Intuitively, it is the EP, not the Landauer cost,
that reflects the ``nitty gritty details'' of the the dynamics of the rate matrix
implementing the computation.) In this subsection
I review recent results establishing precisely how $W$ determines the dependence of EP on $p_0$.

It has long been known how the entropy production \textit{rate},
at a single moment $t$, jointly depends on the rate matrix $W(t)$ and on the 
distribution over states $p_t$. (In fact, those dependencies are given by the expression
in~\erf{eq:ep_def}.) On the other hand, 
until recently nothing was known about how the EP of a discrete time process, evolving over
an extended time interval, depends on the initial distribution over states. 
Initial progress was made in~\cite{kolchinsky2016dependence}, in which the dependence
of EP on the initial distribution was derived for the special case where
$\pi(x_1 \mid x_0)$ is nonzero for all $x_0, x_1$. However, this
restriction on the form of $\pi$ is violated in deterministic computations. 

Motivated by this difficulty,~\cite{wolpert2018exact} extended the earlier work
in~\cite{kolchinsky2016dependence}, to give
the full dependence of EP on the initial distribution for arbitrary $\pi$.
That extended analysis shows that EP can always be written as a sum of two terms. Each of 
those terms depends on $p_0$, as well as on the ``nitty gritty details'' of
the process, embodied in $W(t)$.

The first of those EP terms depends on $p_0$ linearly. By appropriately constructing
the nitty gritty details of the system (e.g., by having the system implementing $\pi$ run
a quasi-static process), it is possible to have this first term equal
zero identically, for all $p_0$. The second of the EP terms instead is
given by a drop in the KL divergence between $p$ and a distribution $q$ that
is specified by the nitty gritty details, during the time interval
$t \in [0, 1]$. For nontrivial distributions $\map$, this term can\textit{not}
be made to equal zero for any distribution $p_0$ that differs from $q_0$.
This is unavoidable EP incurred in running the system, which arises whenever one changes the
a.

To review these recent results,  
recall the definition of islands $c$ and associated distributions $\Delta_c$
from \cref{sec:notation}. Nest make the following definition:
\begin{definition}
\label{def:prior}
For any conditional distribution $\map$ implemented by a CTMC,
and any
island $c \in L(\map)$, the associated \textbf{prior} is
\eq{
    q^c \in  \argmin_{\rr : \supp(r) \in \Delta_c} \EP{\PP}(\rr) \nonumber
}
We write the associated lower bound on EP as
\eq{
    \EPmin{\PP}(c) &:= \min_{\rr : \supp(r) \in \Delta_c} \EP{\PP}(\rr)    \nonumber
}
\end{definition}
It will simplify the exposition to introduce an
arbitrary distribution over islands, $q(c)$, and define
\eq{
q^\PP(x) := \sum_{c \in L(\map)} q(c) q^c(x) \nonumber
}
In~\cite{wolpert2018exact} it is shown that
\eq{
\EP{\PP}(p) = \DDf{p}{q^\PP} - \DDf{\map p}{\map q^\PP}  + \sum_{\mathclap{c \in L(\map)}} p(c) \EPmin{\PP}(c) 
\label{eq:27}
}
(Due to the definition of islands, while the choice of distribution $q(c)$ affects
the precise distribution $q$ inside the two KL divergences, it has no effect on their
difference, and so has no effect on EP; see~\cite{wolpert2018exact}.)

The drop of KL divergences on the RHS of~\erf{eq:27} is called the
the \textbf{mismatch cost} of running the CTMC on the initial distribution $p$, and is written
as $\EF_\PP(p)$.\footnote{In~\cite{kolchinsky2016dependence}, due to a Bayesian interpretation of $q$,
the mismatch cost is instead called the ``dissipation due to incorrect priors''.} 
Given the priors $q^c$, both of these KL divergences in the mismatch cost 
depend only on $p$ and on $\map$. By the data-processing inequality for KL divergence,
mismatch cost is non-negative. It equals zero if $p^c = q^c$ for all $c$, 
or if $\pi$ is a measure-preserving
map, i.e., a permutation of the elements of $X$.

The remaining sum on the RHS of~\erf{eq:27} is called
the \textbf{residual EP} of the CTMC. It is a linear function of $p(c)$,
without any information theoretic character. In addition, it has
no explicit dependence on $\map$. It is (the $p(c)$-weighted average of) the 
minimal EP within each island. $\EPmin{\PP}(c) \ge 0$ for all $c$,
and residual EP equals zero if and only if the process is thermodynamically reversible. 
I will refer to $\EPmin{\PP}(c)$ as the \textbf{residual EP (parameter) vector} of the
process. The ``nitty-gritty'' physics details of 
how the process operates is captured by the residual EP  vector together with the
priors.

Combining \erf{eq:27} with the definitions of EF and of cross entropy establishes the following
set of equivalent ways of expressing the EF:
\begin{proposition}
\label{prop:EF_formula}
The total EF incurred in running a process  that 
applies map $\map$ to an initial distribution $p$ is
\begin{align*}
\WWW_\proc(p) &= \WW(p,\map) + \EF_\PP(p) + \sum_{c \in L(\map)} p(c) \EPmin{\PP}(c) \\
&= [\KKf{p}{q} - \KKf{\map p}{\map q}]  + \sum_c p(c) \EPmin{\PP}(c)
\end{align*}
\label{prop:central_equation_entropy_dynamics}
\end{proposition}
\noindent
Unlike the generalized Landauer's bound, which is an inequality, 
Prop.~\ref{prop:EF_formula} is exact. It holds for both macroscopic and
microscopic systems, whether they are computational devices or not.

I will use the term \textbf{entropic cost} to broadly
refer to entropy flow, entropy production, mismatch cost,
residual entropy, or Landauer cost. Note that the entropic
cost of any computational device is only properly defined if we
have fixed the distribution over possible inputs of the device. 

It is important to realize that we can\textit{not} ignore the residual EP when calculating EF
of real-world computational devices. 
In particular, in real-world computers --- even real-world
quantum computers, presuming they are coupled to input/output devices 
--- a sizable portion of the heat generation
occurs in the wires connecting the devices inside the computer
(often a majority of the heat generation, in fact). However,
wires are designed to simply copy their inputs to their outputs, which is a logically invertible map.
As a result, the Landauer cost of running a wire is zero (to within the accuracy
of the wire's implementing the copy operation with zero error), no matter
what the initial distribution over states of the wire $p_0$ is. For the same reason,
the mismatch cost of any wire is zero. This means that the entire EF incurred by
running any wire is just the residual EP incurred by running that wire.
So in real-world wires, in which $\EPmin{\PP}(c)$ invariably varies with $c$ (i.e., in which
the heat generated by using the wire depends on whether it transmits a 0 or a 1), 
the dependence of EF on the initial distribution $p_0$ must be linear. 
In contrast, for the other devices in a computer (e.g., the digital gates
in the computer), both Landauer cost and mismatch cost can be quite large,
resulting in nonlinear dependencies on the initial distribution.


\begin{example}
It is common in the literature to decompose
the rate matrix into a sum of rate matrices of one or more \textbf{mechanisms} $v$:
\eq{
W_{x;x'}(t) &= \sum_\nu W^v_{x,x'} (t)
\label{eq:W_mult_mech}
}
In such cases one replaces the definitions of
the EF rate and EP rate in~\erf{eq:EF_rate_def},\eqref{eq:ep_def}, 
with the similar definitions,
\eq{
\sum_{x',x'',\nu} W^\nu_{x',x''}(t) p_{x''}(t) 
    \ln \bigg[\frac{W^\nu_{x',x''}}{W^\nu_{x'',x'}}\bigg]
\label{eq:EF_rate_def_multi_mechanisms}
}
and
\eq{
\sum_{x',x'',\nu} W^\nu_{x',x''}(t) p_{x''}(t)
   \ln \bigg[\frac{W^\nu_{x',x''}p_{x''}(t)}{W^\nu_{x'',x'}p_x'(t)}\bigg]
\label{eq:EP_rate_def_multi_mechanisms}
}
respectively. 

When there is more than one mechanism, since 
the log of  a sum is not the same as the sum of a log, these redefined
EF and EP rates differ from the analogous quantities given by plugging 
$\sum_\nu W^v_{x,x'} (t)$ 
into~\erf{eq:EF_rate_def},\eqref{eq:ep_def}. For example,
if we were to evaluate~\erf{eq:EF_rate_def} for this multiple-mechanism $W(t)$,
we would get 
\eq{
\sum_{x',x'',\nu} W^\nu_{x',x''}(t) p_{x''}(t) 
    \ln \bigg[\frac{\sum_{\nu'} W^{\nu'}_{x',x''}}{\sum_{\nu''} W^{\nu''}_{x'',x'}}\bigg]
}
which differs from the expression in~\erf{eq:EF_rate_def_multi_mechanisms}.

Nonetheless, all the results presented above apply
just as well with these redefinitions of EF and EP. In particular, under
these redefinitions the time-derivative
of the entropy still equals the difference between the EP rate and the EF rate, total
EP is still non-negative, and total EF is still a linear function of
the initial distribution. Moreover, that linearity of EF means that
with this redefinition we can still write (total) EP as a sum of the mismatch cost, 
defined in terms of a prior, and a residual EP that is a linear function of the initial
distribution.

By themselves, neither the pair of definitions in~\erf{eq:EF_rate_def},\eqref{eq:ep_def}
nor the pair in~\erf{eq:EF_rate_def_multi_mechanisms},\eqref{eq:EP_rate_def_multi_mechanisms} 
is ``right'' or ``wrong''. Rather, the primary basis for choosing between them
arises when we try to apply the resulting mathematics to analyze 
specific thermodynamic scenarios. The development 
starting from~\erf{eq:EF_rate_def},\eqref{eq:ep_def}, for a single mechanism, can
be interpreted as giving heat flow rates and work rates for the thermodynamic 
scenario of a single heat bath coupled to the system. (See Ex.~\ref{ex:3a} and~\ref{ex:3b} above.) 
However, in many thermodynamic scenarios there are multiple heat baths coupled to the system. 
The standard approach for analyzing these scenarios is to identify each heat bath with a separate
mechanism, so that there is a separate temperature for each mechanism,
$T^\nu$. Typically one then assumes \textbf{local detailed balance} (LDB), meaning that
separately for each mechanism $\nu$, the associated matrix $W^\nu(t)$ 
obeys detailed balance for the
(shared) Hamiltonian $H(t)$ and resultant ($\nu$-specific) Boltzmann distribution defined
in terms of the temperature $T^\nu$, i.e., for all $\nu, x, x', t$,
\eq{
\frac{W^\nu_{x,x'}(t)}{W^\nu_{x',x}(t)} &= e^{[H_{x'}(t) -H_{x}(t)] / T^\nu}
\label{eq:ldb}
}

This allows us to identify
the EF rate in~\erf{eq:EF_rate_def_multi_mechanisms} as the 
rate of heat flow to all of the baths. So the EP rate in~\erf{eq:EP_rate_def_multi_mechanisms} is the rate of 
irreversible gain in entropy that remains after accounting for that EF rate and for the change
in entropy of the system.
See~\cite{van2015ensemble,esposito_three_2010,seifert2012stochastic}.
\label{ex:multiple_baths}
\end{example}

As a final comment, it is important to emphasize that
all of the analysis above assumes that there are no constraints on how
the physical system can implement $\map$. For example, the Landauer cost given in \cref{eq:land_cost_def} and \cref{prop:EF_formula} is the
{unconstrained} minimal amount of EF necessary to implement the conditional distribution $\map$
on any physical system, when there are no restrictions on
the rate matrix underlying the dynamics of that system. However, in practice there will always be \textit{some}
constraints on what rate matrices the engineer of a system can use to implement
a desired logical state dynamics. In particular, the architectures of 
the computational machines defined in \cref{sec:comp_machines} constrain which
variables in a system implementing those machines
are allowed to be directly coupled with one another by the rate matrix.

The minimal amount of EF needed to implement a desired distribution $\map$ if one
is constrained to use a particular computational machine to implement that dynamics
is called the \textbf{machine Landauer cost}. Trivially, since it is the solution
to the same optimization problem that defines Landauer cost, only with
extra constraints imposed, the machine Landauer cost is never
less than the (unconstrained) Landauer cost, i.e. the machine Landauer cost
of some particular computational machine is never less than the value given
in \cref{eq:land_cost_def}. In fact, the machine Landauer cost can
be substantially greater than the unconstrained Landauer cost, as illustrated by
the following simple example. 

\begin{example}
Suppose our computational machine's state space is two bits, $x^1$ and $x^2$, and that
the function $f(x)$ erases both of those bits. Let $p_0(x)$ be the initial
distribution over joint states of the two bits. (As a practical matter,
$p_0$ would be determined by the preferences of the users of the system,
e.g., as given by the frequency counts over a long time interval in which they
repeatedly use the system.)  In this scenario, the unconstrained Landauer cost is  
\eq{
S(p_0(X)) - S(p_1(X)) &= S(p_0(X)) \nonumber \\
    &= S(p_0(X^1)) + S(p_0(X^2 \mid X^1))
}

Now modify this scenario by supposing that we are constrained to 
implement the parallel bit erasure with
two subsystems acting independently of one another, one subsystem acting
on the first bit and one subsystem acting on the second bit. This changes the Landauer
cost to 
\eq{
S(p_0(X^1)) - S(p_1(X^1)) + S(p_0(X^2)) - S(p_1(X^2) &= S(p_0(X^1)) + S(p_0(X^2)) 
}
The gain in Landauer cost due to the constraint --- the machine Landauer
cost --- is $S(p_0(X^2)) - S(p_0(X^2 \mid X^1))$.
This is just the mutual information between the two bits under the initial
distribution $p_0$, which in general is nonzero. 

To understand the implications of this phenomenon, suppose that the parallel bit erasing subsystems are thermodynamically
reversible \textit{when considered by themselves}. It is still the
case that if they are run in parallel as two subsystems
of an overall system, and if
their initial states are statistically correlated, then that overall system is not
thermodynamically reversible. Indeed, if we start with $p_0$, implement the
parallel bit erasure using two thermodynamically reversible bit-erasers, and then run that process 
in reverse, we end up with the distribution $p_0(x^1)p_0(x^1)$ rather than $p_0(x^1, x^2)$. 
\label{ex:double_bit_erasure}
\end{example}

A general analysis of how the architecture of a computational machine 
affects the associated machine Landauer cost can be found in~\cite{wolpert2018exact}.
The special case when the open system in question is an information
ratchet, there is only a single heat bath, and local detailed balance holds, is considered in~\cite{Boyd:2018aa}.  
See also~\cite{riechers_thermo_comp_book_2018,grochow_wolpert_sigact2018}.


It is important to emphasize that all of the results in this section
for mismatch costs and residual EP hold for processes that
implement general stochastic matrices
as well as those that implement single-valued functions, 
if one uses the appropriate generalization of islands. 
Finally, despite the importance of EP to the entropic costs of real world systems,
from now on I
will assume that the residual EP of any island of any device I
consider equals zero. The resultant analysis gives the least possible
EF, which would arise if (as is almost
always the case in the real world) we don't design the prior of a device
to match the actual distribution of its input states, and so must account for its possible
mismatch cost. In addition, this assumption
simplifies the calculations, since we no longer have to consider 
the islands of the processes.

\section{Entropy dynamics of logically reversible circuits}
\label{sec:fredkin}

Our modern understanding
of nonequilibrium statistical physics makes clear that there is no \textit{a priori} relation
between the logical reversibility of a function taking inputs to outputs, $f : X^{IN} \rightarrow X^{OUT}$,
and the thermodynamic reversibility of a system that implements $f$~\cite{sagawa2014thermodynamic,wolpert2016free,wolpert2016correction,wolpert_arxiv_beyond_bit_erasure_2015,esposito2010finite,maroney_absence_2005}. 
This should not be surprising; logical reversibility
is all about maps from the \textit{state} of a system at one time
to its state at another, whereas thermodynamic
reversibility is all about trajectories of \textit{marginal probability distributions}
over the states of the system as the system evolves from one time 
to another (together with properties of any external reservoirs
the system is coupled to, etc.). 
These are completely different kinds of mathematical structures, with the result that 
thermodynamic reversibility need not imply anything about logical reversibility, or vice versa.

However, the pioneering work of Landauer and 
Bennett~\cite{landauer1961irreversibility,benn73,benn82} on the thermodynamics
of computation led to a common misperception that
logical and thermodynamic reversibility are in fact identical.
This has in turn motivated research on ``reversible circuits''. In this research one 
is presented with a conventional circuit $C$ made of
logically \underline{{\textit{ir}}}reversible gates which implements some logically 
irreversible function $f$, and tries to construct a logically \textit{reversible} circuit, $C'$, 
that emulates $C$. The underlying insight is that  we can always do this, by appropriately
wiring together a set of logically reversible gates (e.g., Fredkin gates) to
create a circuit $C'$ that maps any input bits $x^{IN} \in X^{IN}$ to a set of
output bits that contain both $f(x^{IN})$ and a copy of 
$x^{IN}$~\cite{fredkin1982conservative,drechsler2012reversible,perumalla2013introduction,frank2005introduction}.
Tautologically, the entropy of the distribution over the states of this
circuit after the map has completed is identical to the entropy of the initial distribution over states. So the Landauer cost is zero, it would appear. This has led 
to claims in the literature suggesting that by replacing a
conventional logically irreversible circuit with an equivalent logically reversible circuit,
we can reduce the ``thermodynamic cost'' of computing $f(x^{IN})$ to zero. More precisely,
it would appear that we can reduce the total EF expended to zero.

This general line of reasoning should be worrisome. As mentioned, we now know that
we can directly implement any logically irreversible map
$x^{IN} \rightarrow f(x^{IN})$ in a thermodynamically reversible manner. So by running
such a direct implementation of a logically irreversible
$f$ in reverse (which can be done thermodynamically
reversibly), we would extract heat from a heat bath. If we do that,
and then implement $f$ forward using a logically reversible circuit, 
we would return the system to its starting distribution, seemingly having
extracting heat from the heat bath, thereby violating the second law. 

As it turns out, there \textit{are} some thermodynamic advantages to using a logically reversible
circuit rather than an equivalent logically irreversible circuit. However, there are also
some disadvantages to using logically reversible circuits. Moreover, the advantages of
logically reversible circuits
cannot be calculated simply by counting ``the number of bit erasures'' performed by the 
equivalent logically irreversible circuit. In the following two subsections I
elaborate these relative advantages and disadvantages of using reversible circuits, 
in order to illustrate the results presented in the sections above.

Before doing that though, in the remainder of this subsection I present some needed
details concerning logically reversible circuits that are constructed out of logically reversible gates.
One of the properties of logically reversible gates that initially caused problems in
designing circuits out of them is that
running those gates typically produces ``garbage'' bits, to go with the bits that provide the output
of the conventional gate that they emulate. The problem is that these garbage bits
need to be reinitialized after the gate is used, so that the gate can be used again.
Recognizing this problem, \cite{fredkin1982conservative} shows how to
avoid the costs of reinitializing any garbage bits produced by using a reversible
gate in a reversible circuit $C'$, by extending $C'$ with yet more reversible gates (e.g.,
Fredkin gates). The result is an \textbf{extended circuit} that
takes as input a binary sting of input data $x$, along with a binary string of  ``control signals'' $m \in M$,
whose role is to control the operation of the reversible gates in the circuit. The output
of the extended circuit is a binary string of the desired output for input $x^{IN}$, 
$x^{OUT} = f(x^{IN})$, together with a copy 
of $m$, and a copy of $x^{IN}$, which I will write as $x^{IN}_{copy}$. 
So in particular, none of the output garbage bits produced by the individual
gates in the original, unextended circuit of reversible gates still exists by the time
we get to the output bits of the extended circuit.\footnote{More precisely, in one 
popular form of reversible circuits, a map $f : X^{IN} \rightarrow X^{OUT}$ is
implemented in several steps. First, in a ``forward pass'', the circuit made out of reversible gates
sends $(x^{IN}, m, \vec{0}^{GARBAGE}, \vec{0}^{OUT}) \rightarrow (x^{IN}, m, m', f(x^{IN}))$, 
where $m'$ is the set of ``garbage bits'',
$\vec{0}^{OUT}$ is defined as the initialized state of the output bits, and
similarly for $\vec{0}^{GARBAGE}$. After completing this forward pass,
an offboard copy is made of $x^{OUT}$,
i.e., of $f(x^{IN})$. Then the original circuit is run ``in reverse'', sending
$(x^{IN}, m, m', f(x^{IN})) \rightarrow (x^{IN}, m, \vec{0}^{GARBAGE}, \vec{0}^{OUT})$. The end result is a process that transforms the input bit string $x^{IN}$ into the offboard copy
of $f(x^{IN})$, together with a copy of $x^{IN}$ (conventionally stored in the same physical variables that contained
the original version of $x^{IN}$), all while leaving the control bit string $m$ unchanged.}
 
While it removes the problem of erasing the garbage bits, this extension of the original
circuit with more gates does not come for free. In general it requires doubling
the total number of gates (i.e., the circuit's size), doubling the running time of the circuit (i.e., the
circuit's depth), and increasing the number of edges coming out of each gate, by up to a factor
of 3. (In special cases though, these extra cost can be reduced, sometimes substantially.)

\subsection{Reversible circuits compared to all-at-once devices}

In practice, typically we will want to reuse any given circuit many
times, with different inputs each time. Most simply, we can assume that
those inputs are generated by IID sampling a fixed distribution (which
is ultimately determined by the user of the circuit). To ensure that the (average) entropic
costs are independent of how many times the circuit has
been used, we need to reinitialize the output variables of the circuit
after each use. I will refer to a process that does this as the 
\textbf{answer-reinitialization} of that circuit.

Now in general, there are many different ``basis sets'' of allowed gates we can use to
construct a conventional (logically irreversible) 
circuit that computes any given logically irreversible 
function $f$. Indeed, even once we fix a set of allowed gates, in general
there are an infinite number of logically 
irreversible circuits that implement $f$ using that set of gates. 
Due to this, we need to clarify precisely what  ``logically irreversible circuit'' we wish to compare to any given extended circuit that implements the same function $f$ as that circuit. 

One extreme possibility is to compare the extended circuit to a single,
monolithic gate that computes the same function, i.e., to a physical system that directly maps 
$(x^{IN}, \vec{0}^{OUT}) \rightarrow (x^{IN}, f(x^{IN}))$. However,
this map is logically reversible, just like the extended circuit, and
so not of interest for the comparison. 

A second possibility is to compare the extended circuit to a system with a state
space $X$ that directly maps $x \in X \rightarrow f(x) \in X$, without
distinguishing input variables and output variables. Such a map is \textit{not}
logically reversible, but (as mentioned above) can be implemented with
a thermodynamically reversible system, whatever the initial distribution over
$X$. I will refer to such a physical system
that maps $x \rightarrow f(x)$ in a manner that is thermodynamically reversible
for some initial distribution $q(x)$ as an \textbf{all at once} (AO) device,
or AO circuit. (The reason for this terminology is that
the underlying Hamiltonian may need to simultaneously couple all components of $x$ in
order to achieve thermodynamic reversibility). 
If we implement $f$ with an AO device, then the minimal
EF we must expend to calculate $f$ is the drop in entropy of the distribution over $X$ as 
that distribution evolves according to $f$. This drop is
nonzero (assuming $f$ is not logically invertible). This would seem to mean that there
is an advantage to using the equivalent extended circuit rather than the AO device,
since the minimal EF with the extended circuit is zero.

However, we must be careful to compare apples to apples. The number of information-carrying bits
in the
extended circuit after it completes computing $f$ is $\log |X^{IN}| + \log |X^{OUT}| + \log |M|$.
The number of information-carrying bits in the AO device when it completes is 
just $\log |X^{OUT}|$. Therefore the minimal number of bits there must be in the AO device is max$(\log |X^{OUT}|, \log |X^{IN}|)$. So it may be that
the extended circuit and the AO circuit  
implement functions over different spaces, of different sizes. 

This means that the entropic costs of answer-reinitializing the two circuits
(i.e., reinitializing their variables in preparation for the next inputs) will differ.
%
%
In general, the Landauer cost and mismatch cost of {answer-reinitialization}
of an extended circuit will be greater than
the corresponding answer-reinitialization costs of an equivalent AO device. This is for the 
simple reason that the answer-reinitialization of the extended circuit must reinitialize
the bits containing copies of $x$ and $m$, which do not even exist in the AO device. 

To be more quantitative, first, for simplicity, assume that the initial distribution over the bits 
in the extended circuit that encode $m$ is a delta
function. (This would be the case if we do not want the physical circuit to implement 
a different computation from one run to the next,
so only one vector of control signals $m$ is allowed.) This means that
the ending distribution over those bits is also a delta function.
The Landauer cost of reinitializing
those bits is zero, and assuming that we perform the reinitialization using a prior
that equals the delta function over $m$, the mismatch cost is also
zero. So assuming the residual EP of reinitialization those
bits containing a copy of $m$ is zero, we can we can ignore those bits from now on.

To proceed further in our comparison of the entropic costs of the 
answer reinitialization of an AO device
with those of an equivalent extended circuit, we need to specify the detailed dynamics of the
answer-reinitialization process that is applied to the two devices.
Both the AO device and the equivalent extended circuit have a set of output bits
that contain $f(x^{IN})$ that need to be reinitialized, with some associated
entropic costs. In addition though, the extended circuit needs to reinitialize
its ending copy of $x^{IN}$, whereas there is no such requirement of the equivalent AO device.
To explore the consequences of this, 
I now consider several natural models of the answer-reinitialization:

$ $

\noindent 1) In one model, we require that the
answer-reinitialization of the circuit is performed within each output bit $g$ itself, 
separately from all other variables. Define $Fr(\CC)$ to mean an extended circuit 
that computes the same input-output function $f^\CC$
as a conventional circuit $\CC$, and define $AO(\CC)$ similarly. 
Assuming for simplicity that the residual entropy of reinitializing all output bits
is zero,
the EF for the answer-reinitialization of $Fr(\CC)$ using such a bit-by-bit process is
\eq{
\W_{\CC'}(p, q) =  \sum_{g \in V^{\Cout}} \KKf{p_g}{q_g}
}
where $V^{\Cout}$ indicates the set of all bits containing the final values 
of $x^{OUT}$ and $x^{IN}_{copy}$.

Using gate-by-gate answer-reinitialization,
the EF needed to erase the output bits containing $f^\CC(x^{IN})$ is the same
for both $AO(\CC)$ and $Fr(\CC)$. Therefore the additional Landauer cost incurred in answer-reinitialization
due to using  $Fr(\CC)$ rather than $AO(\CC)$ is
the Landauer cost of erasing the output bits in $Fr(\CC)$ that store $x^{IN}_{copy}$,
\eq{
\Delta \SSS_{Fr(\CC), \CC} (p) & \coloneqq  \sum_{v \in V_{IN}} \SSS(p_v)
\label{eq:52}
}
where I write ``$v \in V_{IN}$'' to mean the output bits that contain $x^{IN}_{copy}$,
and $p_v$ to mean the ending marginal distributions over those bits. Similarly,
the difference in mismatch cost is
\eq{
\Delta \DD_{Fr(\CC), \CC} (p, q) & \coloneqq  \sum_{v \in V_{IN}} \DDbase_v(p^v\Vert {q}^v)
}
where $q_v$ refers to a prior used to reinitialize the output bits in .$v \in V_{IN}$.

However, independent of issues of answer-reinitialization,
the Landauer cost of implementing a function using an AO
device that is optimized for an initial distribution $p_{\Cin}$ can be bounded as follows:
\eq{
\SSS(p_{\Cin}) - \SSS(f^\CC p_{\Cin}) &\le \SSS(p_{\Cin})  \nonumber \\
	& \le  \sum_{v \in V_{\Cin}} \SSS_v(p_v) \nonumber \\
	& = \Delta \SSS_{Fr(\CC), \CC}(p)
}
Combining this with \erf{eq:52} shows that under gate-by-gate answer-reinitialization, the \textit{total} 
Landauer cost of implementing a function using an AO device  --- including the costs of
reinitializing the gates containing the value $f^C(x^{IN})$ ---  is
upper-bounded by the \textit{extra} Landauer cost of implementing that same
function with an equivalent extended circuit, i.e., just that portion
of the cost that occurs in answer-reinitializing
the extra output bits of the extended circuit.
This disadvantage of using the extended circuit holds even if the 
equivalent AO device is logically irreversible. So as far as Landauer cost is concerned
there is no reason to consider using an extended circuit
to implement a logically irreversible computation with
this first type of answer-reinitialization.

On the other hand, in some situations, the mismatch
cost of running the AO device will be \emph{greater} 
than the mismatch cost of the answer-reinitialization of $x^{IN}_{copy}$ 
in the equivalent extended circuit. This illustrated in the following example:
\begin{example}
Suppose that the input to the circuit consists of two bits, $a$ and $b$,
where the actual distribution over those bits, $\p$, and prior distribution over
those bits, $q$, are:
\eqn{
\p(x_b) &= \delta(x_b, 0) \\
\p(x_a \mid x_b = 0) &= 1/2 \qquad  \forall x_a \\
q(x_b) &= \delta(x_b, 1) \\
q(x_a \mid x_b = 0) &= 1/2 \qquad  \forall x_a \\ 
q(x_a \mid x_b = 1) &= \delta(x_a, 0)
}

Suppose as well that $f^\CC$ is a many-to-one map. Then plugging in gives
\eq{
\DDf{p_{\Cin}}{q_{\Cin}} - \DDf{f^\CC p_{\Cin}}{f^\CC q_{\Cin}} &=    
    \DDf{p_{\Cin}}{q_{\Cin}}  \nonumber \\
& > \sum_{v \in V_\Cin} \DDf{p_v}{q_v} \nonumber 
}
This sum equals the mismatch cost of the answer-reinitialization of $x^{IN}_{copy}$, which
establishes the claim.
\end{example}

\noindent However, care should be taken in interpreting this result,
since there are subtleties in comparing mismatch costs between
circuits and AO devices, due to the need to compare apples to apples 
(see discussion of this point in~\cite{wolpert2018exact}).

$ $

\noindent 2) A second way we could answer-reinitialize an extended circuit involves
using a system that simultaneously accesses \textit{all} of the output bits
to reinitialize $x^{IN}_{copy}$, including the bits storing $f(x^{IN})$.

To analyze this approach, for simplicity assume there are no restrictions on
how this reinitializing system operates, i.e., that it is an AO device.
The Landauer cost of this type of answer-reinitialization
of $x^{IN}_{copy}$ is just $\SSS(p(X^{IN} \mid X^{OUT})) - \ln[1]$,
since this answer-reinitialization process is a many-to-one map over the state of $x^{IN}_{copy}$. 
Assuming $f^\CC$ is a deterministic map though, by Bayes' theorem
\eq{
\SSS(p(X^{IN} \mid X^{OUT})) &= \SSS(p(X^{IN})) - \SSS(p(X^{OUT}))
}
So in this type of answer-reinitialization, the extra Landauer cost of the 
answer-reinitialization in the extended
circuit that computes $f^\CC$ is identical to the total Landauer
cost of the AO device that computes the same function
$f^\CC$. 
On the other hand, in this type of answer-reinitialization process the mismatch cost of the extended circuit
may be either greater or smaller than that of the AO device, depending
on the associated priors.
%

%

$ $

\noindent 3) A third way we could answer-reinitialize $x^{IN}_{copy}$ in an extended circuit
arises if, after running the circuit, we happened upon a set of initialized external bits, just
lying around, as it were, ready to be exploited. In this case, after running the circuit, we could simply swap
those external bits with $x^{IN}_{copy}$, thereby answer-reinitializing the output bits at zero cost.

Arguably, this is more sleight-of-hand
than a real proposal for how to re-initialize the output bits. Even so, it's
worth pointing out that rather than use those initialized external bits to contain a copy of $x^{IN}_{copy}$,
we could have used them as an information battery, extracting up
to a maximum of $\ktlntwo$ from each one by thermalizing it. So the opportunity cost
in using those external bits to reinitialize the output bits of the extended circuit
rather than use them as a conventional battery is $|V_{IN}| \ktlntwo$. This is an upper
bound on the Landauer cost of implementing the desired computation using an AO device. So again,
as far as Landauer cost is concerned, there is no advantage to using an extended circuit
to implement a logically irreversible computation with this third type of answer-reinitialization.

%

$ $

Summarizing, it is not clear that there is a way to implement a logically irreversible function
with an extended circuit built out of logically reversible gates that
reduces the Landauer cost below the Landauer cost of an equivalent AO device. 
The effect on the mismatch cost of using such a circuit rather than an AO device is 
more nuanced, varying with the priors, the actual distribution, etc.

\subsection{Reversible circuits versus irreversible circuits}

I now extend the analysis from comparing the entropic costs of an extended circuit and those  
of a computationally equivalent {AO device}, to also consider the costs of a
computationally equivalent conventional circuit, built with multiple logically irreversible gates. 
As illustrated below, the entropic costs of the answer-reinitialization of a conventional circuit 
(appropriately modeled) are the same as the entropic costs
of the answer-reinitialization of a computationally equivalent AO device.
So the analysis of the preceding subsection carries over, giving the
relationship between the entropic costs of answer-reinitialization of conventional
circuits and the entropic costs of answer-reinitialization of computationally equivalent extended circuits.
In particular, the minimal EF required to answer-reinitialize a conventional circuit is in general
lower than the minimal EF required to answer-reinitialize a computationally equivalent 
extended circuit. 

Accordingly, in this subsection
I focus instead on comparing the entropic costs of running conventional
circuits, before they undergo any answer-reinitialization,
with the entropic costs of running computationally equivalent extended circuits, before
\textit{they} undergo any answer-reinitialization.
While the full analysis of the entropic costs of running conventional circuits is rather
elaborate~\cite{wolpert2018exact}, some of the essential points can be illustrated with 
the following simple example.

Suppose we have a system that comprises two input bits and two output bits, 
with state space written as
$X = X^{IN}_1 \times X^{IN}_2 \times X^{OUT}_1 \times X^{OUT}_2$. Consider mapping
the input bits to the output bits by running the ``parallel  bit
erasure'' function. Suppose that while doing that we simultaneously reinitialize the
input bits $x^{IN}_1$ and $x^{IN}_2$, in preparation for the next run 
of the system on a new set of inputs.
So assuming both of the output bits are initialized before the process begins to 
the erased value $0$, the state space evolves according to the function 
$f : (x^{IN}_1, x^{IN}_2, 0, 0) \rightarrow (0, 0, 0, 0)$.
 
Consider the following three systems that implement this $f$:
\begin{enumerate}
\item An AO device operating over
$X^{IN}_1 \times X^{IN}_2 \times X^{OUT}_1 \times X^{OUT}_2$ that directly implements $f$;
\item A system that implements $f$ using two bit-erasure gates that are physically
isolated from one another, as
briefly described in~\cref{ex:double_bit_erasure}. Under this model
the system first uses one bit-erasure gate to send $(X^{IN}_1, X^{OUT}_1) = (x^{IN}_1, 0) \rightarrow (0, 0)$,
and then uses a second bit-erasure gate to apply the same 
map to the second pair of bits, $(X^{IN}_2, X^{OUT}_2)$. 

The requirement that the gates be physically isolated means that the rate matrix of the first gate
is only allowed to involve the pair of bits $(X^{IN}_1, X^{OUT}_1)$, i.e., it
is of the form $W_{x^{IN}_1, x^{OUT}_1;(x^{IN}_1)', (x^{OUT}_1)'}(t)$. So
the dynamics of $(x^{IN}_1, x^{OUT}_1)$ is independent of
the values of the other variables, $(x^{IN}_2, x^{OUT}_2)$. Similar restrictions apply to the
rate matrix of the second gate. (So in the language of~\cite{wolpert2018exact}, the two gates each run
a ``solitary process''.) 

\label{enum:2}
\item A system that uses two bit-erasure gates to implement $f$, just as in (\ref{enum:2}), but does \textit{not}
require that those gates run in sequence and that
they be physically isolated. In other words, the rate matrix that
drives the first bit-erasure gate as it updates the variables $(x^{IN}_1, x^{OUT}_1)$ \textit{is} allowed to 
do so based on the values $(x^{IN}_2, x^{OUT}_2)$, and vice-versa. Formally,
this means that both of those rate matrices are of the form
$W_{x^{IN}_1, x^{OUT}_1,x^{IN}_2, x^{OUT}_2;(x^{IN}_1)', (x^{OUT}_1)';(x^{IN}_2)', (x^{OUT}_2)'}(t)$.
\label{enum:3}
\end{enumerate}

\noindent
(\ref{enum:2}) and (\ref{enum:3}) are both physical models of a conventional circuit made out of two gates
that each implement logically-irreversible functions. However, they differ
in whether they only allow physical coupling among the variables in the circuit that
are logically needed for the circuit to compute the desired function (model (\ref{enum:2})), or
instead allow arbitrary coupling, e.g., to reduce entropic costs (model (\ref{enum:3})).

The Landauer cost of the first model, (1), is the minimal EF needed to implement the parallel  bit erasure
with a single monolithic gate, 
\eq{
&S(p_0(X^{IN}_1, X^{IN}_2,
X^{OUT}_1, X^{OUT}_2)) - S(\hat{f}_{1,2} \; p_0(X^{IN}_1, X^{IN}_2, X^{OUT}_1, X^{OUT}_2)) \nonumber \\
&\qquad = S(p_0(X^{IN}_1, X^{IN}_2, X^{OUT}_1, X^{OUT}_2))  \nonumber \\
&\qquad = S(p_0(X^{IN}_1, X^{IN}_2))
}
where $\hat{f}_{1,2}$ is the conditional distribution implementing the parallel  bit erasure, 
so that $\hat{f}_{1,2} \, p_0(x)$ is the ending distribution, which is 
a delta function centered at $(0,0,0,0)$.

Next, assume that both of the gates in (\ref{enum:2}) are thermodynamically reversible,
if they are considered in isolation, separately from any other systems (and if they are run
on the appropriate initial distributions). In other words, \textit{considered
in isolation from any other systems}, they are both AO devices.
So the minimal EF needed to run the first of those gates is 
\eq{S(p_0(X^{IN}_1) - S(\hat{f}_1 \; p_0(X^{IN}_1, X^{OUT}_1)) = S(p_0(X^{IN}_1)
}
Similarly, the minimal EF needed to run the second gate is $S(p_0(X^{IN}_2)$.

Combining, we see that the difference between \{the minimal EF needed to
run a conventional circuit constructed as in model (\ref{enum:2})\} and \{the minimal EF
needed to run a computationally equivalent AO device (model (1))\} is
\eq{
S(p_0(X^{IN}_1) + S(p_0(X^{IN}_2) - S(p_0(X^{IN}_1, X^{IN}_2))
}
This is just the initial mutual information between $X^{IN}_1$ and $X^{IN}_2$. So
the minimal EF needed to run model (\ref{enum:2}) will exceed the minimal EF needed to
run model (1) whenever $X^{IN}_1$ and $X^{IN}_2$
are statistically coupled under the initial distribution, $p_0$.
(See \cref{ex:double_bit_erasure}.)

On the other hand, because of the increased flexibility in their rate matrices, we can
assume that the bit-erasure gates in the circuit defined in (\ref{enum:3}) each achieve zero EP 
\textit{even when considered as systems operating over the full set of four bits}. So each of those bit-erasure gates
is thermodynamically reversible even when considered in the context of the full system.
As a result, running the circuit defined in model (\ref{enum:3}) requires the same minimal
EF as running an AO device. So in general, the minimal EF needed to
run to the conventional circuit defined in model (\ref{enum:3}) is less than the minimal EF needed to
run to the conventional circuit defined in model (\ref{enum:2}).
This increase in the minimal EF of running a circuit as modeled in (\ref{enum:2}) compared to
a circuit as modeled in (\ref{enum:3}) is precisely the difference between
machine Landauer cost and (unconstrained) Landauer cost, discussed at the end of \cref{sec:entropy_dynamics}.
(A detailed analysis of this difference in costs, involving 
explicit rate matrices, can be found in the appendices in~\cite{wolpert2018exact}.)

Next, note that we only need to reinitialize the two output bits after running 
any of the three models (1), (\ref{enum:2}) and (\ref{enum:3}). In contrast, the 
extended circuit that runs $f$ needs to reinitialize its input bits as well, in preparation 
for receiving a new set of inputs for a next run of the system. So 
models (1), (\ref{enum:2}) and (\ref{enum:3}) have lower minimal EF for answer-reinitialization than
the computationally equivalent extended circuit, in general.

Summarizing, the minimal total EF 
(including both the EF needed to run the system and to 
answer-reinitialize it) that is needed by model (\ref{enum:2}) exceeds
the minimal total EF needed by either the equivalent AO device (model (1))
or the equivalent conventional circuit as defined by model (\ref{enum:3}), with the difference
equal to the mutual information of the inputs bits under $p_0$.
In turn, the minimal EFs to (only) run either model (1) or model (\ref{enum:3})
exceeds the minimal EF needed to run an equivalent extended circuit, by
$S(p_0(X^{IN}_1, X^{IN}_2))$. However, the minimal \textit{total} EF of
the models (1) and (\ref{enum:3}) will in general be no greater than the minimal
total EF of the extended circuit, and may be smaller (depending on the details of the answer-reinitialization process in the extended circuit).


On the other hand, as a purely practical matter, constructing a conventional circuit
as in (\ref{enum:3}) for circuits substantially larger than parallel bit-erasures may be quite challenging;
to do so requires that identify all sets of variables that are \textit{statistically} coupled,
at any stage of running the circuit,
and make sure that our gates are designed to \textit{physically} couple those variables.
There are no such difficulties with constructed an extended circuit.
Another advantage of an extended circuit is that {no matter what
the true distribution $p_0$ is}, an extended circuit has zero mismatch cost,
since there is no drop of KL divergence between $p_0$ and \textit{any} $q_0$
under a logically reversible dynamics. In contrast, all three models (1) - (\ref{enum:3}) 
can have nonzero mismatch cost, in general.

As yet another point of comparison, an extended circuit will often have far more wires
than an equivalent conventional circuit. And as mentioned above, the residual EP generated
in wires is one of the major sources of EF in modern digital gates. So even 
in a situation where a conventional circuit has nonzero mismatch cost, 
when the EF generated in the wires is taken into account, there may be no 
disadvantage to using that conventional circuit rather a computationally equivalent
extended circuit.

Clearly there is a rich relationship between the detailed wiring diagram of a conventional
logically irreversible circuit, the procedure for
answer-reinitializing the outputs of a computationally equivalent extended circuit, 
the distribution over the input bits of those circuits, and how the aggregate
entropic costs of those two circuits compare. Precisely delineating this relationship
is a topic for future research.

\begin{acknowledgments}
\emph{Acknowledgements} ---
I would like to thank Josh Grochow and Peter Stadler for helpful discussion,
and thank the Santa Fe Institute for helping to support this research. This paper was made possible through 
Grant No. CHE-1648973 from the U.S. National Science Foundation and Grant No.
FQXi-RFP-1622 from the FQXi foundation. 
The opinions expressed in this paper are those of the author and do not necessarily 
reflect the view of the SFI, the National Science Foundation, or FQXi.
\end{acknowledgments}

\bibliographystyle{aipnum4-1}

%

\newcommand{\arXiv}[2]{\href{http://arxiv.org/abs/#1}{arXiv:#1 #2}}

\end{document}